\documentclass[aip,jap,amsmath,amssymb,reprint,groupedaddress]{revtex4-1}
\usepackage{graphicx}
\usepackage{bm}
\usepackage[colorlinks=true,linkcolor=blue]{hyperref}

\pdfoutput=1

\begin{document}

\preprint{AIP/123-QED}

\title{Direct Imaging of Radio-Frequency Modes via Traveling Wave Magnetic Resonance Imaging}

\author{A. Tonyushkin$^1$}
\thanks{Present address: Physics Department, University of Massachusetts Boston, Boston, MA}
\email{alexey.tonyushkin@umb.edu}
\author{D. K. Deelchand$^2$}
\author{P.-F. Van de Moortele$^2$}
\author{G. Adriany$^2$}
\author{A. Kiruluta$^{1,3}$}
\affiliation{$^1$Massachusetts General Hospital, Harvard Medical School, Boston, MA}
\affiliation{$^2$Center for Magnetic Resonance Research, University of Minnesota, Minneapolis, MN}
\affiliation{$^3$Physics Department, Harvard University, Cambridge, MA}


\begin{abstract}
We demonstrate an experimental method for direct 2D and 3D imaging of magnetic radio-frequency (rf) field distribution in metal-dielectric structures based on
traveling wave (TW) magnetic resonance imaging (MRI) at ultra-high field (\textgreater 7T). 
The typical apparatus would include an ultra-high field
whole body or small bore MRI scanner, waveguide elements filled with MRI active dielectrics with predefined electric and magnetic
properties, and TW rf transmit-receive probes. We validated the technique by obtaining TW magnetic-resonance (MR) images of the magnetic field distribution
of the rf modes of circular waveguide filled with deionized water in a 16.4 T small-bore MRI scanner and compared the MR images
with numerical simulations. Our MRI technique opens up a practical non-perturbed way of imaging of previously inaccessible rf field distribution of 
modes inside of various shapes metal waveguides with inserted dielectric objects, including waveguide mode converters and transformers. 
\end{abstract}
\maketitle

\section{Introduction}
Imaging of the rf magnetic field distribution is important for various applications that involve designing waveguide elements,
mode splitters, and mode converters, where specific rf (microwave) field profile is desirable
for engineering high-power rf propagation over the long distances, microwave heating in various research and commercial applications \cite{Vinogradov1991, Aleksandrov1992, Choe2013}.
Most of the measurements of the static magnetic field are done by utilizing various probes and consecutive mapping the field point by point. 
Electromagnetic field patterns currently can be imaged from surface radiometry by using thermographic pattern analyzers \cite{Balageas1993, Vernieres2011}. 
Previously the dielectric waveguides were studied in optical band and corresponding electromagnetic mode patterns were observed in the optical fibers \cite{Snitzer1961, Doerr2008}. However modes imaging in rf band is still technologically challenging.
Recent advents of ultra-high field MRI make it possible to
propagate resonant rf waves in overmoded waveguide regime inside MRI scanners \cite{BrunnerMRM2011}. The propagating rf transverse magnetic field component becomes a spin excitation field for an MRI active medium thus imprinting a map of the transverse $H_1$ field distribution (in MRI a flux density
$B_1$ is commonly used instead of H-field) into a sample. Such an image of the field map can be obtained in real time provided
that the waveguide is physically compatible with the high DC magnetic field inside the bore of MRI scanner. 
A complimentary way to obtain an rf field distribution is by computer simulation using various electromagnetic field solvers \cite{Collins2011} that
require a lot of computational resources and multiple solvers for solution validation. In our work we developed and demonstrated a TW MRI technique to directly image rf modes inside a circular waveguide filled with MR active high-permittivity dielectrics.

In MRI, $H_1$ field maps of near-field coils at the resonance frequency of the MR scanner are obtained using a uniform dielectric sample in a process called rf shimming for optimization of custom-built coils \cite{Insko1993, Vaughan1994}. Therefore, the modes of such volume near-field coils are well characterized through both computer simulations and MRI. 
In recent work, MRI images of the modes of cylindrical dielectric resonators placed in high-field MR scanner were reported \cite{Wen1996, Webb2012}. Although, conceptually similar to our method the latter relies on dielectric resonances of isolated cylinders with exact dimensions that correspond to specific dielectric resonances.

At high field MRI ($> 4$~T) the propagating waves effects inside dielectrics become important \cite{Foo1992, Kiruluta2007, Tonyu2012} and can be used for imaging provided that the cut-off requirements for the mode propagation is fulfilled. Traveling wave (TW) MRI is a far-field imaging technique that relies on rf mode propagation in a cylindrical magnet bore utilized as an rf waveguide. This approach was originally experimentally demonstrated at 7 T \cite{Brunner2009} and at clinically relevant field of 3 T \cite{Muller2012, Vazquez2013}. However, the main challenge for the TW propagation in most conventional MRI scanners and spectrometers is the relatively small bore radius-to-wavelength ratio ($a/\lambda$) even at ultra-high field strengths. 
The cut-off condition in the cylindrical bore is given by the dispersion relationship: $ h_{mn}=\sqrt{k^2-(\chi'_{mn}/a)^2} $, 
where {\em h} is the propagating vector, {\em m, n} are integer numbers that are related to azimuthal and radial field variations, respectively, {\em k} is the k-vector of free propagating wave inside a dielectric (with $\varepsilon, \mu$), {\em a} is the radius of the guide, $\chi'_{mn}$ is the {\em n}-th root of the derivative of the Bessel function $J_m$ of the first order {\em m} (m=0, 1, 2É). So the cut-off frequency for the lowest TE mode is
 $ f_c=\chi' _{mn}/(2\pi a\sqrt{\mu \varepsilon})$ and the critical wavelength of a mode is $\lambda_c=c/f_c$.
For example, in a 7 T human-size MRI system with a metal guide fitted to the bore (maximum diameter of 60 cm) at a resonant frequency of 298 MHz (free space wavelength $\lambda=100.6$~cm), only a single  mode ($TE_{11}$) will propagate in an unloaded bore. 
Although most of TW MRI were done in a large bore MRI scanner at 7 T, rf imaging might be impractical there due to relatively small field of view (FOV $\sim$ 50cm) to the rf wavelength ratio ({\it e.g.}, $\lambda =30$~cm inside a dielectric with $\varepsilon_r$=80). At high polarizing $B_0$ fields (high resonance frequency) the rf
wavelength inside a waveguide with a dielectric insert that has a high $\varepsilon_r$ becomes small enough to facilitate propagating modes of such a hybrid waveguide. In our work we incorporated a dielectric waveguide that fits into FOV of a 16.4 T
(698 MHz) MRI scanner with a metal screen formed by the gradient insert itself and devised the transverse rf magnetic field distribution of such waveguide from the intensity pattern of MR images.

\section{Method and Theory}
\begin{figure}[tb!]
\includegraphics[width=3.2 in]{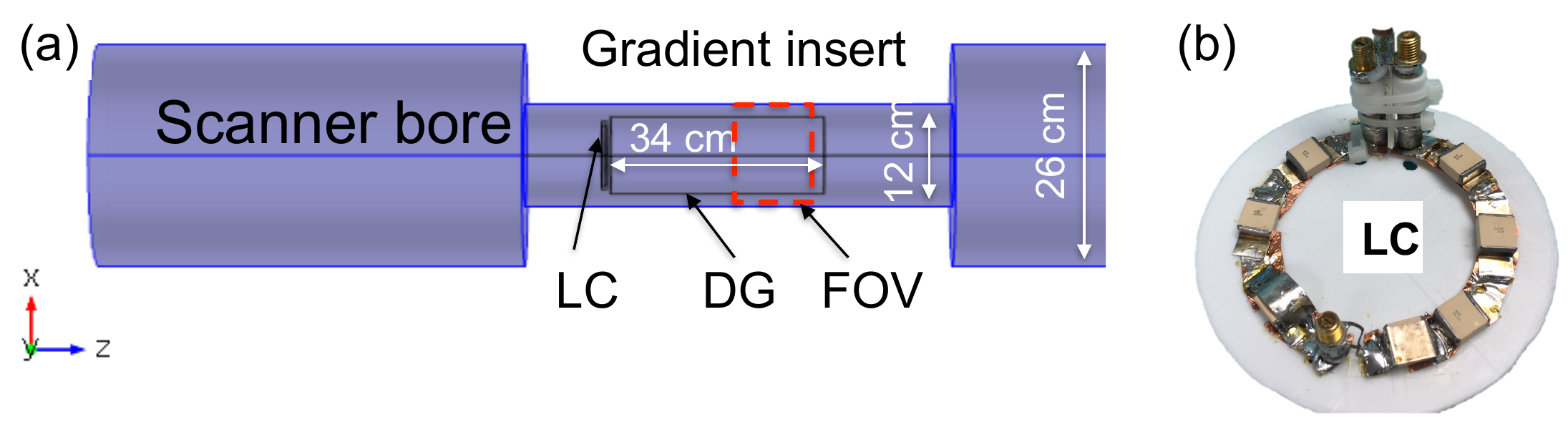}
\caption{Setup: a) an MRI scanner comprised by the metal bore with an image encoding gradient insert, a dielectric guide (DG): acrylic tube filled with Di-water,  FOV=field-of-view, and b) a loop-coil (LC) oriented in axial plane. The rf mode propagates several wavelengths to the imaging FOV (dashed rectangle); its location with respect to the end of the tube can be varied. }
\label{setup}
\end{figure}
The complete solutions for the components of {\bf H} field for TE (TM) waves inside a hollow circular waveguide are given by analytical expressions below
~\cite{Jackson1998}:
\begin{eqnarray}
\label{TEfield}
 \nonumber 
H_z=c_m J_m \left( \chi_{mn}\, r/a \right) \cos (m\phi) e^{-ih_{mn} z},\\ 
H_r=\frac{-i a h_{mn} c_m}{\chi^2_{mn}} J'_m \left( \chi_{mn}\, r/a \right) \cos (m\phi) e^{-ih_{mn} z} , \\ \nonumber
H_{\phi}=\frac{i m a^2 h_{mn} c_m}{\chi^2_{mn} r} J_m \left( \chi_{mn}\, r/a \right) \sin (m\phi) e^{-ih_{mn} z},
\end{eqnarray}
where $c_m$ is a constant. The field for TM modes can be derived from the Helmholtz equation by setting $H_z=0$ and the dispersion
relationship is modified by substituting $\chi'_{mn}$  for $\chi_{mn}$ -- the n-th zero of the Bessel function.
Knowing the {\bf H} field, the respective components of E-field can be derived from the impedance condition: ${\bf H}=[\hat{z} \times {\bf E} ]/\eta_{TE}$, where $\eta$ is the transverse impedance of the mode. In MRI, $H_z$ component does not contribute to MR signal, but it can be important to consider for mode excitations. 
In our effectively oversized waveguide with a high-permittivity dielectric insert, in principle, there are eight supported TW modes that can be excited and imaged as listed in Table~\ref{table}. The number of the modes in the fixed geometry setup can be further controlled by changing the permittivity of the dielectric inside the tube.
\begin{table}[b!]
\caption{Propagating modes in the oversized circular waveguide with dielectric insert: mode index, critical wavelength $\lambda_c$, and propagating period $\Lambda=1/h$.}
\label{table}
\begin{tabular*}{\linewidth}{@{\extracolsep{\fill}} cc|cccccccc@{}}
\hline \hline
Mode && $TE_{11}$ & $TM_{01}$ & $TE_{21}$ & $TE_{01}$, $TM_{11}$ & $TM_{21}$ & $TE_{12}$ & $TM_{02}$ \\ 
 \hline 
$\lambda_c (cm)$ && 15.4 & 11.8 & 9.3 &  7.4 & 5.5 & 5.3 & 5.1  \\
$\Lambda (cm)$ && 5 & 5.3 & 5.6 &  6.3 & 9.8 & 11.3 & 13.8  \\
\hline \hline 
\end{tabular*}
\end{table}
In addition to hollow waveguide modes, the high reactance dielectric boundary conditions allow hybrid modes EH (or HE), which consist of linear combinations of TE and TM field components and therefore $E_z, H_z \neq 0$. A combined system of a hollow metal guide that is partially filled with a concentric dielectric provides a radially anisotropic medium, and generally is not solved analytically \cite{TonyISMRM11}. 
Here, we consider a practical case of a quasi-uniform dielectric filling, where the dielectric-metal air gap is small compared to 
the wavelength and the $H_1$ field is approximated by a field of a uniform waveguide.

The rf modes could in principle be excited with a set of properly positioned elementary probes such as electric and magnetic dipoles. Due to wide
availability of inductive loop-coils (magnetic dipoles) for near-field MRI we use these coils as probes to inductively couple into the dielectric guide. 
To predict the excitation modes these probes have to be inserted inside the dielectric, in practice however, we place them into the fields extrema of the specific mode outside of the dielectric \cite{arxiv}. We consider two distinct cases: loop placed concentric (x-y plane) with the waveguide and orthogonal to the front edge of the waveguide (y-z plane). In the first case the magnetic field ${\bf H} \| \hat{\bf z}$ couples into TE mode (Eq.(\ref{TEfield})). In the second case 
${\bf H} \perp \hat{\bf z} $ and the probe predominantly excites lowest $TE_{11}$ mode. 

In general, the relationship between the MR image intensity and magnetic field map is non-trivial and specially designed MRI sequences exist to provide magnitude and phase magnetic field maps \cite{Yarnykh2007, Pohmanna2013}. Such sequences usually target spatial profiles of $H^+_1$ and $H^-_1$  separately, which are the right- and left-hand polarized components of $H_1$ that correspond to transmit and receive sensitivity profiles of the rf coil respectively. For our application we are interested in obtaining actual propagating $H_1$ field map from the MR image. 
It turned out that the MR signal intensity could provide a qualitative magnitude map of the $H_1$ field for some important cases.
For a gradient-recalled echo imaging sequence the signal intensity in each voxel is given by \cite{Hoult2000}
\begin{equation}
\label{signal1}
s \propto \rho \sin (V|H^{+}_1|\gamma \tau) |(H^{-}_1)^*| ,
\end{equation}
where we neglected relaxation effects; $\rho$ is spin density distribution, $\gamma$ is the gyromagnetic ratio, $\tau$ is the rf excitation pulse duration, $V$ is a dimensionless normalization factor proportional to coil driving voltage. Furthermore, in a small flip angle regime, which typically holds for TW MRI, the Eq.~(\ref{signal1}) can be rewritten as $s \propto |H^{+}_1| |(H^{-}_1)^*|$. 
Next we use expressions for linearly driven coil \cite{Hoult2000}:
\begin{equation}
\label{signal2}
H^{+}_1=\frac{1}{2} (H_{1x}+i H_{1y}) , 
H^{-}_1=\frac{1}{2} (H_{1x}-i H_{1y})^* .
\end{equation}
Substituting Eq.~(\ref{signal2}) into $s$ we obtain
\begin{equation}
\label{signal3}
s \propto \frac{1}{4} |H_{1x}^2+ H_{1y}^2| .
\end{equation}
We further notice that the phase of the complex single mode TW field components (see Eq.(\ref{TEfield})) is the spatial TW phase along z-axis, 
$-ihz$, so we can write Eq.~(\ref{signal3}) as
\begin{equation}
\label{signal4}
s \propto \frac{1}{4} |(\tilde{H}_{1x}^2+ \tilde{H}_{1y}^2) e^{-i2(hz-\phi)} |=\frac{1}{4} |{H_1}|^2 ,
\end{equation}
where $\phi$ is a fixed phase ($\pi /2$ for TE modes), $\tilde{H}_{1x,y}$ are magnitudes of the respective field components with no {\it z} dependence. Notice, however, that the reflected wave gives rise to a standing wave pattern along z-axis and $s \propto cos(2hz)$.
So, from the signal intensity distributions of the two orthogonal ({\it i.e.}, coronal and axial) slices one can obtain a qualitative map of the magnitude of the magnetic field distribution. Eq.~(\ref{signal4}) is extended to a standing wave regime that includes a mixture of modes. However, Eq.~(\ref{signal4}) is not in general valid for the pure TW regime with multiple modes propagation and $H_1$ MR mapping sequences are required to obtain the magnetic field distribution.

\section{Experimental results and simulations}
We performed proof of principle experiment on a 16.4 T horizontal-bore magnet interfaced to an Agilent DirectDrive console (Agilent Technologies, CA) with the resonance frequency of 698 MHz ($\lambda$=43 cm).  The magnet bore diameter is 26 cm, which is further reduced with a 12 cm gradient
coil insert capable of reaching 1000 mT/m in 150 $\mu$s (Resonance Research, Inc., MA). The typical setup of a model
waveguide with inductive coupling (Fig.~\ref{setup}) consists of a dielectric waveguide made of an acrylic tube (L=34 cm, outer diameter 10 cm, inner diameter 9 cm) and filled with deionized water or
saline ($\varepsilon_r$=80, $\sigma=5\times10^{-4}$~S/m or $\sigma=0.6$~S/m respectively) and a simple transmit-receive loop-coil (D=7 cm) placed at the
edge of the tube. The respective modified wavelength in the dielectric is $\lambda_{\epsilon}=$4.9~cm. The orientation of the loop with respect to the
edge of the tube (axial or longitudinal planes) defines the specific single or multiple rf modes excitations in the dielectric tube. 
The MR images were obtained with a gradient-recalled echo sequence: FOV=20x20~cm, TR/TE=100/2~ms, flip angle set at 60$^{\circ}$, 10 (coronal and axial) slices with slice thickness of 5~mm, encoding matrix=128x128, and bandwidth=89kHz. 

 \begin{figure}[t!]
 \includegraphics[width=3.2 in]{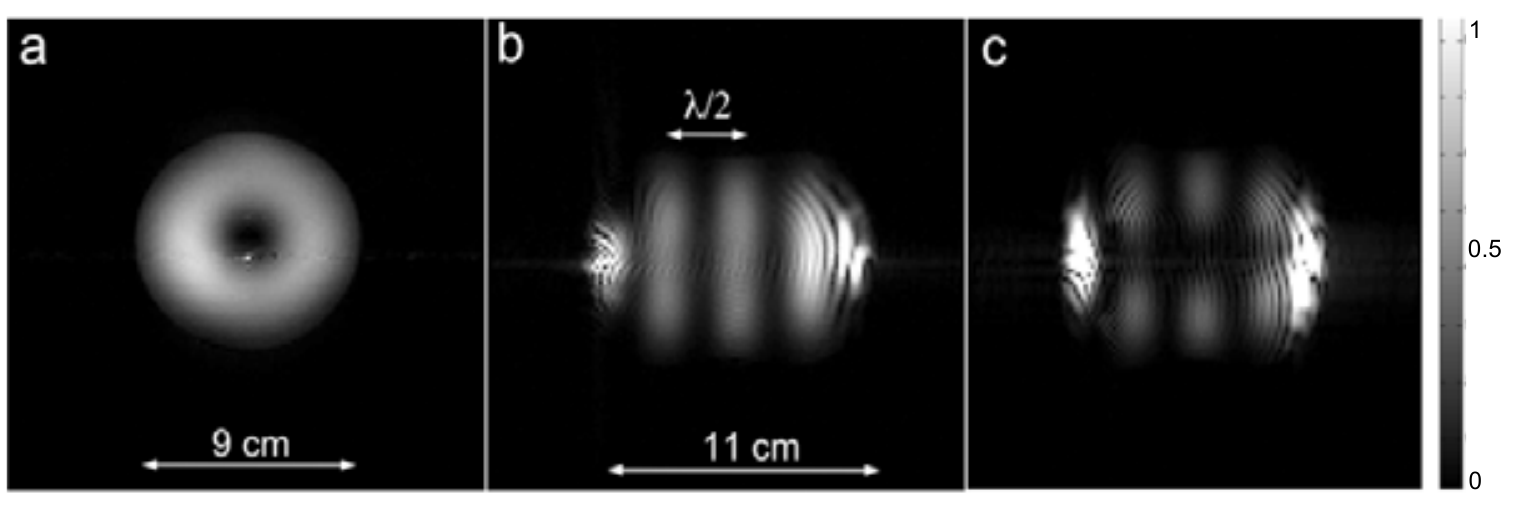}
 \caption{MR images (intensity in arbitrary units) of Di-water tube: axial (a), coronal off-axis (b) and on-axis (c). Dark spot in the center (a) and horizontal strip (c) is the null of $TE_{01}$ mode. Vertical stripes (b, c) are interference fringes with period $\lambda/2 \approx 3$~cm. (Curved bright spots on the edges (b, c) and ringing are artifacts from gradient edges.)}
 \label{TE01}
 \end{figure}
 \begin{figure}[t!]
 \includegraphics[width=3 in]{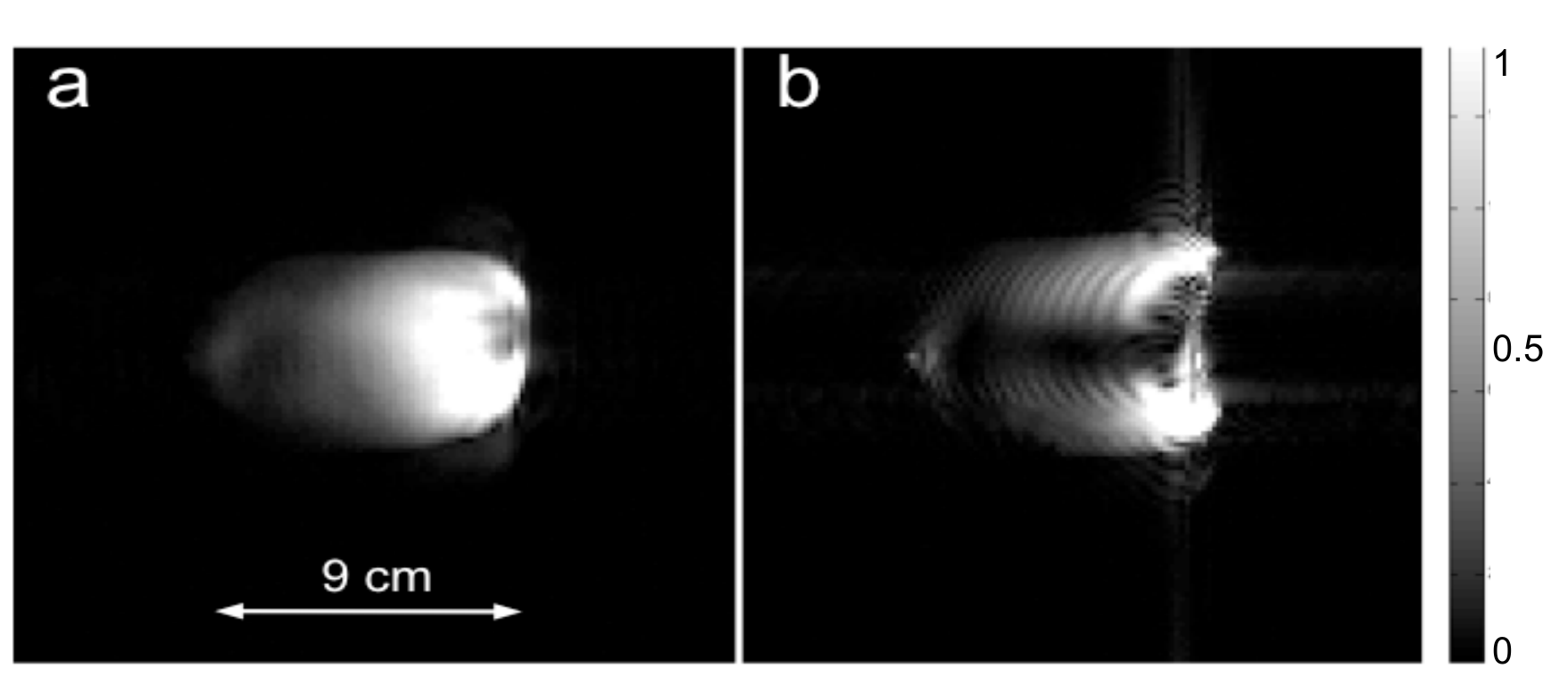}
 \caption{No interference effect: MR images (intensity in arbitrary units) of saline phantom, coronal off-axis (a) and on-axis (b). Saline conductivity strongly attenuates the reflected wave resulting in no interference fringes in this medium (the probe is located on the right side). Here, the gradient-related edge artifacts are amplified with respect to lower intensity decaying signal.}
 \label{saline}
 \end{figure}
In our experimental study we used two excitation methods by a single loop-coil that was oriented in two orthogonal planes: axial and coronal (longitudinal). 
Figure~\ref{TE01} shows MR images obtained for axial coil orientation in a sample of non-conductive dielectric (deionized (Di) water). 
Since the far edge of the dielectric tube represents unmatched boundary we expect a reflected TW thus creating an interference pattern in the FOV
as shown in Fig.~\ref{TE01}. For the actual magnetic field imaging purposes obtaining the interference fringes is complimentary for characterizations of the mode by measuring its period $\Lambda$. On the other hand, to eliminate back reflection and interference fringes we filled the tube with conductive dielectric (water) in otherwise same unmatched boundary setup. The resulted MR image is showing a signal attenuated along z-axis (Fig.~\ref{saline}). In the latter setup the exact period of the mode could be derived from the signal phase plotted vs z coordinate. 
 \begin{figure}[tb!]
 \includegraphics[width=3.2 in]{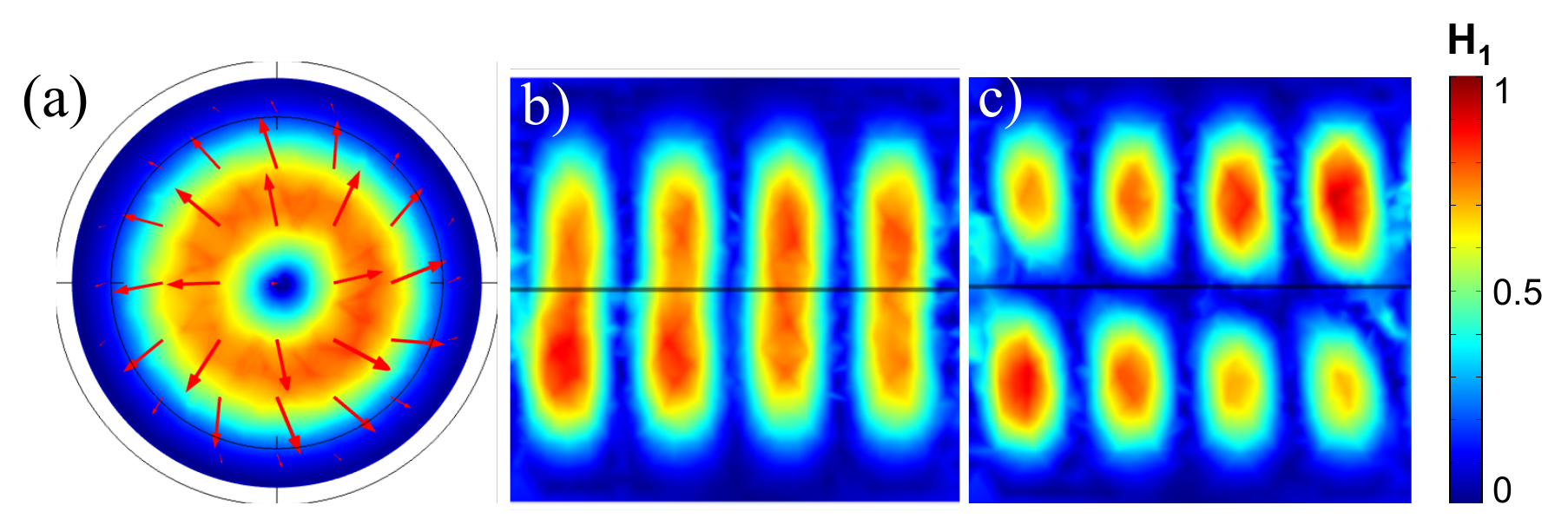}
 \caption{Simulations of $H_1$ field of dominated mode $TE_{01}$ in a partially filled dielectric waveguide: a) axial, b) coronal off- and c) on-axis slices, matching the intensity pattern in Fig.~\ref{TE01}. Arrows indicate $\bf H_1$ vectors, the colorbar shows normalized magnitude of $\bf H_1$. }
 \label{simu}
 \end{figure}

To validate obtained MR images and characterize magnetic field patterns, we carried out simulations of the modes inside the screened dielectric guide using finite elements method (FEM) solver COMSOL Multiphysics (Burlington, MA). The results of the full-wave eigenmode $H_1$ simulations of the same setup as used in Fig.~\ref{TE01} are shown in Fig.~\ref{simu}. The axial cross section view of the simulations shows arrows that depict direction of $\bf H_1$ and correspond to $TE_{01}$ mode. Notice the extended view of the field pattern that covers otherwise MR invisible air all the way to the metal (gradient) boundaries. 
The simulations of $H_1$ field profile show excellent agreement with experimental MRI images.

\begin{figure}[tb!]
 \includegraphics[width=3.2 in]{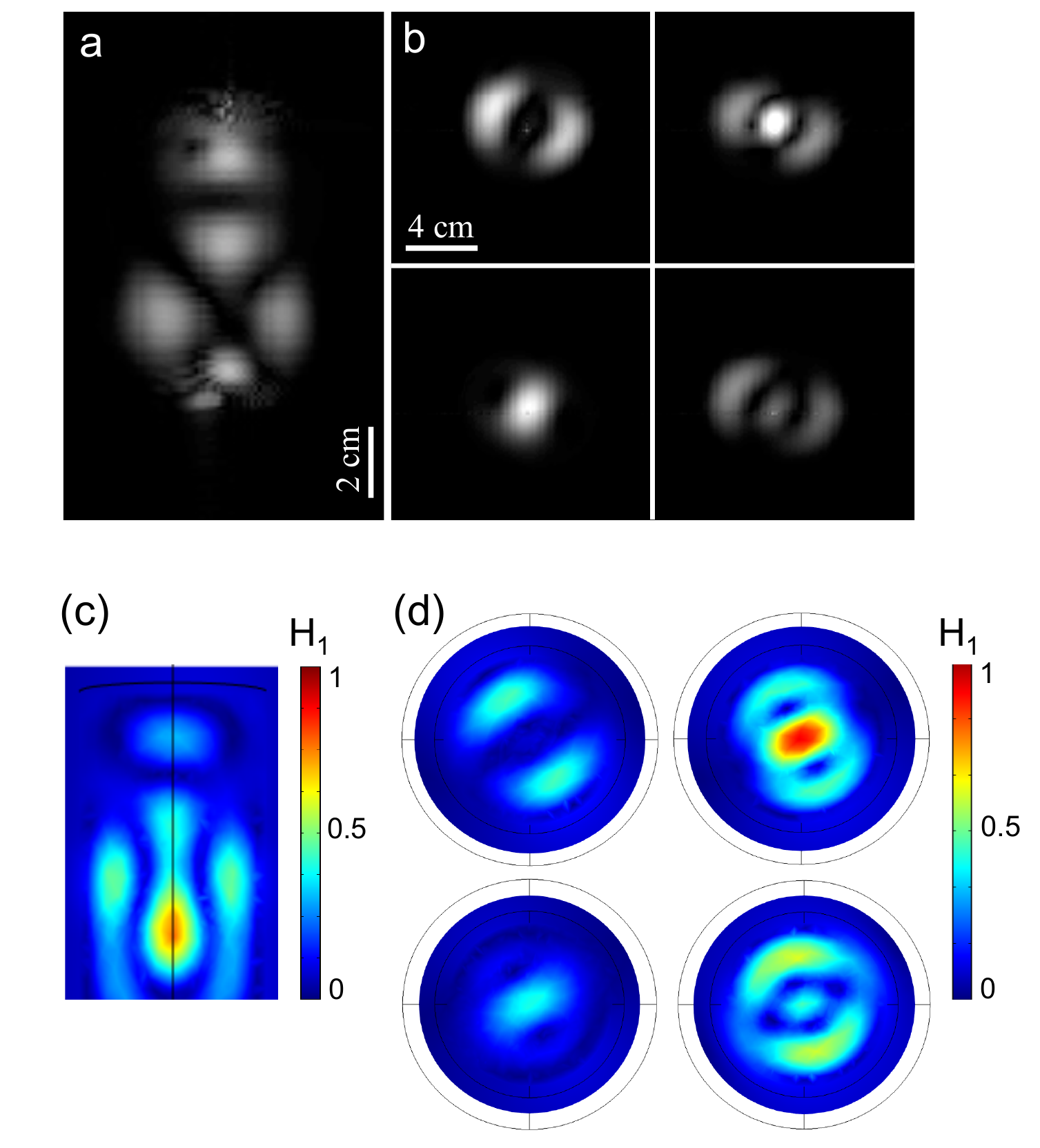}
 \caption{Experimental MR images of Di-water tube with excitation of mixture of modes by the loop-coil oriented in coronal plane: coronal (a) and four axial slices (b). (Here, the intensity scale is arbitrary.)
Simulations of $H_1$ field of mixture of modes in a partially filled dielectric waveguide: c) coronal and d) four axial slices of mixture of modes ($TM_{11}$ and $TE_{11}$); the colorbars show normalized magnitude of $\bf H_1$.}
 \label{simu2}
 \end{figure}

In another imaging example we used an alternative configuration of the loop coil that was positioned in orthogonal plane with 
$H_1 \perp \hat{\bf z}$ and a dielectric tube filled with Di-water. The experimental results are shown in Fig.~\ref{simu2}(a,b). Similarly to the first imaging example we obtained a longitudinal standing wave pattern (Fig.~\ref{simu2}(a)) but with simultaneous excitation of a mixture of modes. The axial cross sections at various position along z (Fig.~\ref{simu2}(b)) show different field patterns implying a mixture of at least two distinctive modes. 
The corresponding simulations of the $H_1$ standing field pattern (Fig.~\ref{simu2}(c,d)) matches the MR images and implies two basic modes: $TE_{11}$ and $TM_{11}$.  
The nulls on z-axis (coronal slices) correspond to the standing waves of modes due to reflection from the far end of the dielectric guide. 
Overall, the imaging experiment produced similar MR pattern to simulations of a relatively complicated field pattern. In the experimental images (coronal slices in Figs.~\ref{TE01},\ref{saline},\ref{simu2}(a)) the high intensity star-like edge brightening and image distortions correspond to the artifacts due to spatially unencoded volume wrap-up of the dielectric guide that extends beyond the gradient allowed FOV and non-linear regions of the gradient coil respectively. In order to eliminate these artifacts one could utilize two compartment dielectric guide with MR non-visible dielectric (ceramic or $D_2 O$) and limit FOV within linear region of the gradient insert.

\section{Conclusions}
In conclusion, we demonstrated the experimental method for direct imaging of magnetic rf field distribution in metal-dielectric waveguides based on TW MRI at ultra-high field. The magnetic field that is generated in MR active medium by the rf pulse facilitates imaging of the modes of circular waveguide and opens up a practical way of non-perturbed imaging of the rf field distribution inside various hybrid waveguides, including waveguide mode converters and transformers. 
We showed theoretically that TW excitation allows qualitative mapping of MR images to the magnetic rf field pattern and experimentally validated the technique by obtaining TW MR images of the magnetic field distribution of the rf modes of circular waveguide filled with deionized water in a 16.4 T small-bore MRI scanner and compared the MR images with numerical simulations.  We found that modern systems with ultra-high magnetic fields ($\geq$ 7 T) are well-suited for the rf field imaging. Our method could be the only direct way to visualize the rf field in the UHF band inside the waveguide structures apart from the numerical simulations. 

\begin{acknowledgments}
We would like to acknowledge help from Dr. M. Garwood from the Center for Magnetic Resonance Research (CMRR). This work was partially supported by grants NIH P41 EB015894 and NIH S10 RR025031. 
\end{acknowledgments}

\bibliographystyle{model1-num-names}

\begin{thebibliography}{10}

\bibitem{Vinogradov1991}
D. V. Vinogradov, G. G. Denisov,
Int. J. of Infrared and Millimeter Waves 12, 131-140 (1991).

\bibitem{Aleksandrov1992}
N. L. Aleksandrov, A. V. Chirkov, G. G. Denisov, D. V. Vinogradov, W. Kasparek, J. Pretterebner, D. Wagner,
Int. J. of Infrared and Millimeter Waves 13, 1369-1385 (1992).

\bibitem{Choe2013}
M. S. Choe , K. H. Kim, EunMi Choi, 
J. Electromag. Waves and Appl. 27, 2221-2238 (2013).

\bibitem{Balageas1993}
D. Balageas, P. Levesque, and A. Deom, 
Proc. SPIE, 1933, 274-285 (1993).

\bibitem{Vernieres2011}
J. Vernieres, J.-F. Bobo, D. Prost, F. Issac, F. Boust, 
IEEE Trans. Magn. 47, 2184 (2011).

\bibitem{Snitzer1961}
E. Snitzer, H. Osterberg, 
J. Opt. Soc. Amer., 51, 499 (1961).

\bibitem{Doerr2008}
C. R. Doerr, H. Kogelnik,
J. Lightwave Tech., 26, 1176 (2008).

\bibitem{BrunnerMRM2011}
D. O. Brunner, J. Paska, J. Froehlich,  K. P. Pruessmann, 
Magn. Reson. Med. 66, 290 (2011).

\bibitem{Collins2011}
C. M. Collins, Z. Wang, 
Magn. Reson. Med. 65, 1470 (2011).

\bibitem{Insko1993}  
E.K. Insko, L. Bolinger, J. Magn. Reson. A 103, 82 (1993).
%
\bibitem{Vaughan1994}
J. T. Vaughan, H. P. Hetherington, J. O. Otu, J. W. Pan, G. M. Pohost,
Magn. Reson. Med. 32, 206 (1994).

\bibitem{Wen1996}
H. Wen, F. A. Jaffer, T. J. Denison, S. Duewell, A. S. Chesnick, R. S. Balaban,
J. Magn. Reson. B. 110, 117 (1996).
%
\bibitem{Webb2012}
A.G. Webb, 
J. Magn. Reson. 216, 107 (2012).

\bibitem{Foo1992}
T. K. Foo, C. E. Hayes, Y. M. Kang,
Magn. Reson. Med. 23, 287 (1992).
%
\bibitem{Kiruluta2007}
A. Kiruluta, J. Phys. D: Appl. Phys.  40, 3043 (2007).

\bibitem{Tonyu2012}
A. A. Tonyushkin, J. A. Muniz, S. C. Grant, and A. M. Kiruluta, Proc. Intl. Soc. Mag. Reson. Med. 20, 2693 (2012).

\bibitem{Brunner2009}
D. O. Brunner, N. De Zanche, J. Froehlich, J. Paska, K. P. Pruessmann, 
Nature 994-998 (2009).

\bibitem{Muller2012}
S. Alt, M. Muller, R. Umathum, A. Bolz, P. Bachert, W. Semmler, and M. Bock, 
Magn. Reson. Med. 67, 1173 (2012).

\bibitem{Vazquez2013}
F. Vazquez, R. Martin, O. Marrufo, and A. O. Rodriguez, 
J. Appl. Phys. 114, 064906 (2013). 

\bibitem{Jackson1998}
J. Jackson, Classical Electrodynamics, 3rd Ed, Wiley (1998).

\bibitem{arxiv} 	
A. A. Tonyushkin, J. A. Muniz, S. C. Grant, and A. M. Kiruluta, arXiv:1409.1965 [physics.med-ph] (2014).

\bibitem{TonyISMRM11} 
A. Tonyushkin, A. M. Kiruluta, Proc. Intl. Soc. Mag. Reson. Med. 19, 1903 (2011).

\bibitem{Yarnykh2007}
V. L. Yarnykh,
Magn. Reson. Med. 57, 192 (2007).
%
\bibitem{Pohmanna2013}
R. Pohmanna, K. Scheffler, NMR Biomed. 26, 265Ð275 (2013).

\bibitem{Hoult2000}
D. I. Hoult, 
Conc. Magn. Reson. 12Ð4, 173 (2000).

\end{thebibliography}

\end{document}